\documentclass[aps,twocolumn,showpacs,preprintnumbers,prb]{revtex4}
\usepackage{graphicx}
\usepackage{dcolumn}
\usepackage{bm}

\begin{document}
\title{Surface-plasmon polaritons in a lattice of metal cylinders}
\author{ J. M. Pitarke$^{1,2}$, J. E. Inglesfield$^{3}$, and 
N. Giannakis$^4$}
\affiliation{
$^1$CIC nanoGUNE Consolider, Mikeletegi Pasealekua 56, E-20009 Donostia, Basque
Country\\
$^2$Materia Kondentsatuaren Fisika Saila, UPV/EHU, and Unidad F\'\i
sica Materiales
CSIC-UPV/EHU,\\ 644 Posta kutxatila, E-48080
Bilbo, Basque Country\\
$^3$School of Physics and Astronomy,
Cardiff University, Cardiff, CF24 3AA, UK\\ $^4$ Department of
Electronic Engineering, Queen Mary, University of London, London, E1 4NS, UK}

\date\today

\begin{abstract}
Plasmon modes of a two-dimensional lattice of long conducting circular
wires are investigated by using an embedding technique to solve
Maxwell's equations rigorously. The frequency-dependent density of states
is calculated for various values of the wave vector and the filling
fraction. At low filling fractions, collective modes are all found to
accumulate at the surface-plasmon frequency $\omega_p/\sqrt{2}$,
$\omega_p$ being the bulk plasmon frequency. As the filling fraction
increases, the interference between the electromagnetic fields
generated by localized surface-plasmon polaritons leads to the
presence of {\it new} resonances, whose frequency strongly depends on the
interparticle separation. For touching wires, a number of multipole
resonances fill the spectral range between dipole resonances, as
occurs in the case of a three-dimensional packing of metal spheres.
\end{abstract}


\maketitle

\section{Introduction}

Over the last few years, a great deal of attention has been devoted to
the investigation of metallo-dielectric photonic crystals, partially
stimulated by the appearance of remarkable optical properties
associated with the excitation of surface-plasmon
polaritons.\cite{pendry2,ebbesen2,halas,barnes,maier,ozsc06} These
are self-sustained collective excitations corresponding to fluctuations of the
surface electron density.\cite{raether,rop}

To understand the properties of surface-plasmon polaritons, and their
possible role in the optical properties of artificially structured
metallo-dielectric metamaterials, we need the photonic band structure
of lattices of metallic inclusions with frequency-dependent dielectric
functions. Lattices of metallic rods,\cite{rods}
cylinders,\cite{maradudin,cylinders,takunori,efros} and
spheres\cite{spheres1,spheres2,spheres3,spheres4} have all been
studied up to now. Three-dimensional (3D) face-centred cubic lattices
composed of metal spheres, their plasmon modes, and interaction with
light were studied by Yannopapas {\it et al.},\cite{spheres3} in the
low filling fraction regime, and wide bands of multipole resonances
were later found to occur as the filling fraction
increases.\cite{spheres4} Flat bands of localized surface plasmons
were also found to occur in a two-dimensional (2D) periodic
arrangement of long metallic cylinders.\cite{takunori} The impact of
localized multipole plasmons on the optical absorption and the energy
loss of 2D and 3D metallo-dielectric crystals was addressed in
Refs.~\onlinecite{pitarke1,pitarke2,javi1}, by looking at the
imaginary part of the so-called effective dielectric function of the
composite.

In this paper, we consider the \emph{evolution} of plasmon bands in a 2D
square lattice made up of long metallic cylinders embedded in a
homogeneous medium, as the cylinder size increases from the non-interacting
limit (small radius compared to the lattice constant) to touching cylinders
where interactions play a key role. We shall consider electromagnetic modes
with the wavevector perpendicular to the cylinders, in
which case the modes fall into two categories: $E$- (or $s$-)
polarization, in which the electric field is parallel to the cylinder
axis, and $H$- (or $p$-) polarization, in which the electric field lies
in the 2D plane. We shall focus our attention on the plasmonic regime, where
the radius of the cylinders is much smaller than the wavelength of the
interacting electromagnetic radiation.  

At long wavelengths, the behaviour
for $E$-polarization is the same as that of a homogeneous metal with a reduced
plasmon frequency given by $\sqrt{f}\omega_p$,\cite{maradudin,pitarke1} where
$f$ is the
filling fraction and $\omega_p$ the bulk plasmon frequency of the metal
cylinders. Conversely, for $H$-polarization surface plasmons can be excited,
and due to the interference between their electromagnetic fields their
frequencies vary strongly with the cylinder separation, which suggests that
their frequency can be {\it tuned} by changing the separation. Here we shall
show that for this polarization and in the limit of touching spheres
multipole resonances fill the spectral range between dipole resonances, as
occurs in the case of 3D lattices of metal spheres.

\section{Theory}

We consider a square array of infinitely long metal
cylinders embedded in a homogeneous medium of dielectric constant
$\epsilon_0$; the metal of the cylinders is described by the
free-electron Drude dielectric function\cite{ashcroft}
\begin{equation}\label{drude}
\epsilon(\omega)=1-{\omega_p^2\over\omega(\omega+i/\tau)},
\end{equation} 
where $\tau$ represents the inelastic scattering time of bulk plasmons. In this
section, we describe our theory, starting off with general considerations and
going on to discuss the embedding method which we use to calculate the
dispersion of the electromagnetic modes.

\subsection{General considerations}

In the long-wavelength limit, in which the radius of the cylinders is much
smaller than the wavelength of the interacting electromagnetic radiation, we
should be able to replace our composite material by an effective homogeneous
medium of dielectric function $\epsilon_{\mbox{\tiny eff}}(\omega)$, such that
electromagnetic modes propagate with frequency $\omega$ given as a
function of wavevector $k$ by
\begin{equation}\label{general}
\omega=kc/\sqrt{\epsilon_{\mbox{\tiny eff}}(\omega)},
\end{equation}
where $c$ is the speed of light. Due to the anisotropy of the material, the
effective dielectric function $\epsilon_{\mbox{\tiny eff}}(\omega)$ depends on
the polarization of the propagating electromagnetic waves.  

\subsubsection{$E$-polarization}
In the case of electromagnetic waves with the electric field polarized
along the cylinders ($s$-polarization), we know from elementary
electromagnetism\cite{jackson} that the electric field is continuous
across the interface. This suggests that in the long-wavelength limit
the composite will behave as a homogeneous medium with the dielectric
function of this effective medium given by the weighted average of the
dielectric functions of the constituents,\cite{bergman,noter}
\begin{equation}\label{epol1}
\epsilon_{\mbox{\tiny eff}}(\omega)=f\epsilon(\omega)+(1-f)\epsilon_0.
\end{equation}
Here, $f$ is the volume fraction of the cylindrical
inclusions. Taking the Drude dielectric function, Eq.~(\ref{drude}),
and cylinders in vacuum ($\epsilon_0=1$), Eq.~(\ref{epol1}) yields
\begin{equation}\label{epol2}
\epsilon_{\mbox{\tiny eff}}(\omega)=1-f{\omega_p^2\over\omega(\omega+i/\tau)}.
\end{equation}
This shows that for this polarization the optical response of the
composite material is expected to be that of free electrons in a
homogeneous electron gas, but with the {\it reduced} plasma frequency
$\sqrt{f}\omega_p$.

\subsubsection{$H$-polarization}
For electromagnetic waves with the electric field polarized normal to
the cylinders, the electric field is strongly modified by the presence
of the interfaces.

In the case of a two-component isotropic system composed of {\it identical}
inclusions of dielectric function $\epsilon(\omega)$ in a host medium of
dielectric constant $\epsilon_0$, the polarization ${\bf P}$ can be easily
obtained as follows
\begin{equation}\label{one}
{\bf P}=f{\epsilon-1\over 4\pi}{\bf E}_{\mbox{\tiny in}}+
(1-f){\epsilon_0-1\over 4\pi}{\bf E}_{\mbox{\tiny out}},
\end{equation}
where ${\bf E}_{\mbox{\tiny in}}$ and ${\bf E}_{\mbox{\tiny out}}$ represent
the average electric field inside and outside the inclusions, respectively.
Assuming that our composite material can be replaced by an effective
homogeneous medium of dielectric function
$\epsilon_{\mbox{\tiny eff}}(\omega)$, one can also write
\begin{equation}\label{two}
{\bf P}={\epsilon_{\mbox{\tiny eff}}-1\over 4\pi}{\bf E},
\end{equation}
where
${\bf E}$ is the macroscopic electric field averaged over the composite:
\begin{equation}\label{hp11}
{\bf E}=f{\bf E}_{\mbox{\tiny in}}+(1-f){\bf E}_{\mbox{\tiny out}}.
\end{equation}
Eqs.~(\ref{one})-(\ref{hp11}) yield the following relation:
\begin{equation}\label{hp1}
(\epsilon_{\mbox{\tiny eff}}-\epsilon_0)\,{\bf E}=
f(\epsilon-\epsilon_0)\,{\bf E}_{\mbox{\tiny in}}.
\end{equation}

In the case of a {\it single} 2D circular inclusion embedded in an otherwise
homogeneous medium, an elementary analysis\cite{jackson} shows that the
electric field ${\bf E}_{\mbox{\tiny in}}$ in the interior of the inclusion is
\begin{equation}\label{hp2}
{\bf E}_{\mbox{\tiny in}}={u\over u-1/2}\,{\bf E},
\end{equation} 
where
\begin{equation}\label{hp3}
u=\left[1-\epsilon/\epsilon_0\right]^{-1}.
\end{equation}
Introducing Eqs.~(\ref{hp2}) and (\ref{hp3}) into Eq.~(\ref{hp1}) yields
\begin{equation}\label{mg0}
\epsilon_{\mbox{\tiny eff}}(\omega)=\epsilon_0\left[1-f{1\over u-1/2}\right].
\end{equation}
This equation, which coincides with Eq.~(27) of Ref.~\onlinecite{efros},
describes for a Drude metal the surface-plasmon mode
$\omega_s=\omega_p/\sqrt{1+\epsilon_0}$ of a {\it single} cylinder. 

The interaction among circular inclusions in a host medium can be
introduced approximately in the framework of the Maxwell-Garnett (MG)
approximation.\cite{bohren} The basic assumption of this approach is that the
average electric field ${\bf E}_{\mbox{\tiny in}}$ within an inclusion
located in a system of identical inclusions is related to the average
field ${\bf E}_{\mbox{\tiny out}}$ in the medium {\it outside} as in
the case of a single {\it isolated} (noninteracting) inclusion,
thereby only dipole interactions being taken into account. Hence, in
this approach the electric field ${\bf E}_{\mbox{\tiny in}}$ is taken
to be of the form of Eq.~(\ref{hp2}) but with the macroscopic electric
field ${\bf E}$ replaced by the electric field ${\bf E}_{\mbox{\tiny
out}}$ outside, which together with Eqs.~(\ref{hp11}) and (\ref{hp1})
yields:\cite{noter}
\begin{equation}\label{mg}
\epsilon_{\mbox{\tiny eff}}(\omega)=\epsilon_0\left[1-f{1\over u-m}\right],
\end{equation} 
with $m=(1-f)/2$. In the dilute ($f\to 0$) limit, $m=1/2$ and the
Maxwell-Garnett Eq.~(\ref{mg}) yields Eq.~(\ref{mg0}).

As optical absorption by a composite is dictated by the poles of
$\epsilon_{\mbox{\tiny eff}}$, an inspection of Eq.~(\ref{mg}) shows that for
Drude cylinders in vacuum the MG approximation predicts optical
absorption to occur at $\sqrt{(1-f)/2}\,\omega_p$, which in the dilute
($f\to 0$) limit yields the surface-plasmon frequency $\omega_p/\sqrt{2}$.

Conversely, the energy loss of swift charged particles is dictated by the poles
of the effective inverse dielectric function
$\epsilon_{\mbox{\tiny eff}}^{-1}(\omega)$. From Eq.~(\ref{mg}), one finds: 
\begin{equation}\label{mg2}
\epsilon_{\mbox{\tiny eff}}^{-1}(\omega)=\epsilon_0^{-1}
\left[1+f{1\over u-n}\right],
\end{equation}
with $n=(1+f)/2$. Equation (\ref{mg2}) shows that for Drude cylinders in vacuum
the MG approximation predicts the excitation of surface plasmons at
$\sqrt{(1+f)/2}\,\omega_p$, which in the dilute
($f\to 0$) limit yields, as in the case of optical absorption, the
surface-plasmon frequency $\omega_p/\sqrt{2}$.

\subsection{Embedding technique}

In the absence of free charges or currents, Maxwell's equations reduce
to the eigenvalue equation
\begin{equation}
\nabla\times\nabla\times\mathbf{E}=\epsilon(\mathbf{r})\frac{\omega^2}
{c^2}\mathbf{E},   \label{max1}
\end{equation}
where $\epsilon(\mathbf{r})$ is the spatially varying dielectric
function, and we assume a magnetic permeability $\mu=1$ everywhere.
To solve this equation, we use the embedding technique described in
Ref.~\onlinecite{ingles1} and employed in Ref.~\onlinecite{spheres4} to
investigate plasmon bands in 3D periodic arrangements of metal
spheres. In this approach, the metallic cylinders are replaced by an
embedding potential over their surfaces, and the electromagnetic field
in between is expanded in a given basis set. With the embedding method one has,
of course, complete freedom in the choice of this basis set; we use a
plane-wave basis set, which represents a natural choice for periodic systems.
Because the plane-wave basis is used only in the vacuum between the cylinders,
convergence is very rapid.

We consider an electromagnetic wave normally incident on the
structure, i.e., with the wave vector taken in the plane perpendicular
to the axes of the cylinders, which we take to be the
$z$-direction. With this choice, the cylinder problem simplifies
considerably compared with the spheres, because the vector
wave-equation can be written in scalar form. In the case of
$E$-polarization, the problem reduces to solving the wave-equation for
$E_z$. The other case is $H$-polarization, with the electric field
perpendicular to the cylinders, in which case we have a scalar
wave-equation for $H_z$. As usual when solving Maxwell's equations,
the boundary conditions on the fields are that the surface-parallel
components of $\mathbf{E}$ and $\mathbf{H}$ are continuous across the
surfaces of the cylinders. This is taken care of by the embedding
potential.

\subsubsection{$E$-polarization}

For $E$-polarization, Eq.~(\ref{max1}) becomes the scalar wave equation,
\begin{equation} \label{max2}
-\nabla^2 E_z=\epsilon(\mathbf{r})\frac{\omega^2}{c^2} E_z. 
\end{equation}
In the embedding method, we derive a variational expression for the
eigenvalue $\omega^2/c^2$ in terms of a trial electric field
$\mathcal{E}_z$ defined only in the region between the cylinders,
region $I$,
\begin{equation} \label{var}
\frac{\omega^2}{c^2}=
\frac{\int_I d\mathbf{r}\nabla \mathcal{E}^{*}_z\cdot\nabla 
\mathcal{E}_z-\int_S d\mathbf{r}_S\int_S d\mathbf{r}'_S 
\mathcal{E}^{*}_z\left(\Sigma-\omega_0^2
\frac{\partial\Sigma}{\partial\omega_0^2}\right)\mathcal{E}_z}
{\epsilon_0\int_I d\mathbf{r}\mathcal{E}^{*}_z\cdot \mathcal{E}_z+c^2\int_S 
d\mathbf{r}_S\int_S d\mathbf{r}'_S\mathcal{E}^{*}_z\frac{\partial
\Sigma}{\partial\omega_0^2}\mathcal{E}_z}.
\end{equation}
The contribution from each cylinder is replaced by the double integral
over its surface, containing the frequency-dependent embedding
potential $\Sigma(\mathbf{r}_S,\mathbf{r}'_S;\omega_0^2/c^2)$,
evaluated at a trial frequency $\omega_0$. The terms involving the
derivatives of the embedding potential originally arise from volume
integrals through the cylinders, but provide a first-order correction
so that $\Sigma$ is evaluated at the estimated frequency $\omega$
rather than the trial frequency $\omega_0$.

The embedding potential is defined in terms of the exact solution
$E_z$ of the wave equation inside the cylinder, at frequency
$\omega_0$, which matches on to the trial solution $\mathcal{E}_z$
over the surface of the cylinder,
\begin{equation} \label{embdef}
\frac{\partial E_z(\mathbf{r}_S)}{\partial n_S}=
\int_S d\mathbf{r}'_S\Sigma(\mathbf{r}_S,\mathbf{r}'_S)\mathcal{E}_z
(\mathbf{r}'_S)
\end{equation}
-- it gives the exact normal derivative corresponding to the trial
surface amplitude.

To solve Eq.~(\ref{var}), the trial function is expanded in terms of basis
functions $F_i$, 
\begin{equation} \label{pw}
\mathcal{E}_z(\mathbf{r})=\sum_i e_iF_i(\mathbf{r}),\;\;
F_i(\mathbf{r})=\exp(i\mathbf{k}_i\cdot\mathbf{r}).
\end{equation}
For a 2D periodic lattice of cylinders, suitable basis functions are
plane waves, where the wave vector $\mathbf{k}_i$ is given by
$\mathbf{k}_i=\mathbf{k}+\mathbf{g}_i$, with $\mathbf{k}$ the Bloch
wave vector (two-dimensional, in the plane perpendicular to the
cylinders) and $\mathbf{g}_i$ is a 2D reciprocal lattice
vector. Substituting into Eq.~(\ref{var}) and finding the stationary
values with respect to the coefficients $e_i$ gives the matrix
eigenvalue equation
\begin{equation}\label{eigen}
Ae=\frac{\omega^2}{c^2}Be,
\end{equation}
where the $A$ and $B$ matrices are given by
\begin{equation} \label{amat}
A_{ij}=\int_I d\mathbf{r}\nabla F^*_i\cdot\nabla F_j
-\int_S \!d\mathbf{r}_S\int_S \!d\mathbf{r}'_S F^*_i
\left(\Sigma-\omega_0^2\frac{\partial\Sigma}{\partial\omega_0^2}
\right)F_j
\end{equation}
and
\begin{equation}  \label{bmat}
B_{ij}=\epsilon_0\int_I d\mathbf{r}F^*_i\cdot F_j+c^2\int_S 
d\mathbf{r}_S\int_S d\mathbf{r}'_S F^*_i\frac{\partial
\Sigma}{\partial\omega_0^2}F_j.
\end{equation}
The integrals in region $I$, over the 2D unit cell
excluding the cylinder, are easy to evaluate,
\begin{equation}
\int_I d\mathbf{r}\nabla F^*_i\cdot\nabla F_j=\left\{
\begin{array}{c}k_i^2(A-\pi r^2),\;i=j\\-2\pi r\mathbf{k}_i\cdot
\mathbf{k}_j\frac{J_1(r|\mathbf{k}_i-\mathbf{k}_j|)}
{|\mathbf{k}_i-\mathbf{k}_j|},\;i\ne j
\end{array}\right.
\end{equation}
and 
\begin{equation}
\int_I d\mathbf{r}F^*_i\cdot F_j=\left\{
\begin{array}{c}(A-\pi r^2),\;i=j\\-2\pi r\frac{J_1(r|\mathbf{k}_i
-\mathbf{k}_j|)}{|\mathbf{k}_i-\mathbf{k}_j|},\;i\ne j
\end{array}\right.
\end{equation}
-- $A$ is the area of the unit cell, and $r$ is the cylinder
radius. 

From a multipole expansion of the solution of the scalar wave equation in
cylindrical harmonics and matching onto the trial wave function on the surface
of the cylinder, Eq.~(\ref{embdef}) yields the following expression for the
matrix element of the embedding potential:
\begin{eqnarray}
\int_S \!d\mathbf{r}_S\int_S \!d\mathbf{r}'_S F^*_i\Sigma F_j&=&
-2\pi r\kappa\sum_m^{m_{\mbox{\tiny max}}}\exp im(\phi_i-\phi_j)\nonumber \\
&\times&\frac{J_m(k_i r)J_m(k_j r)}
{J_m(\kappa r)}J_m'(\kappa r).
\end{eqnarray}
Here, $J_m(x)$ represent Bessel Functions of the first
kind,\cite{abra} $\phi_i$ is the angle of basis wave vector
$\mathbf{k}_i$, and $\kappa$ represents the magnitude of the (complex)
wave vector of the solution of Maxwell's equations inside the cylinder
at trial frequency $\omega_0$,
$\kappa=\sqrt{\epsilon}\,\omega_0/c$. For this polarization,
convergence is rapid with respect to the maximum harmonic,
$m_{\mbox{\tiny max}}$, in the expansion.

The complex dielectric function of the lossy metal cylinders,
Eq.~(\ref{drude}), means that the frequencies of solutions of Maxwell's
equations are broadened, and rather than solving the eigenvalue
equation, Eq.~(\ref{eigen}), we find the corresponding Green function
$\Gamma(\mathbf{r},\mathbf{r}';\lambda)$ given in region $I$ by
\begin{equation}\label{gf1}
\Gamma(\mathbf{r},\mathbf{r}';\lambda)=\sum_{ij}\Gamma_{ij}
(\lambda)F^*_i(\mathbf{r})F_j(\mathbf{r}'),
\end{equation}
with
\begin{equation}\label{gf2}
\sum_k(A_{ik}-\lambda B_{ik})\Gamma_{kj}(\lambda)=\delta_{ij}.
\end{equation}
As we know the frequency at which the Green function is evaluated, the
embedding potential is evaluated at this frequency, and the
frequency-derivative terms in $A$ and $B$ cancel out in
Eq.~(\ref{gf2}).

The Green function is related to the spectral density of the electric
field,
\begin{equation}\label{dos}
n(\mathbf{r},\omega)=\sum_i \epsilon(\mathbf{r}) 
E^*_{z,i}(\mathbf{r})E_{z,i}(\mathbf{r})
\delta(\omega-\omega_i),
\end{equation}
which multiplied by $\delta\omega$ gives the electric-field intensity
at point $\mathbf{r}$ in this frequency range. As the Green function
can be expressed in terms of the eigenmodes and eigenfrequencies
satisfying Eq.~(\ref{max2}),
\begin{equation}\label{green}
\Gamma(\mathbf{r},\mathbf{r}';\lambda)=\sum_i\frac{E^*_{z,i}(\mathbf{r})
E_{z,i}(\mathbf{r}')}{\omega^2_i/c^2-\lambda},
\end{equation}
we see that the spectral density is given by
\begin{equation}
n(\mathbf{r},\omega)=\frac{2\omega}{\pi c^2}\epsilon(\mathbf{r})
\mbox{Im}\Gamma(\mathbf{r},\mathbf{r};\omega^2/c^2+i\delta).
\end{equation}
Integrating $n$ through region $I$ gives 
\begin{equation}
n_I(\omega)=\frac{2\omega\epsilon_0}{\pi c^2}\sum_{ij}\mbox{Im}
\Gamma_{ij}\int_I d\mathbf{r}F^*_i\cdot F_j
\end{equation}
-- $n_I$ is the quantity we calculate and plot.

\subsubsection{$H$-polarization}

For $H$-polarization, we solve the scalar equation for the
$z$-component of the magnetic field,
\begin{equation} \label{max3}
-\nabla^2 H_z=\epsilon(\mathbf{r})\frac{\omega^2}{c^2} H_z, 
\end{equation}
and then the requirement that the surface-parallel components of
$\mathbf{H}$ and $\mathbf{E}$ are continuous reduces to the continuity
of $H_z$ and $\frac{1}{\epsilon}\partial H_z/\partial n_s$ across the
surface of the cylinders. The magnetic variational principle,
analogous to Eq.~(\ref{var}), is given by
\begin{equation}\label{varm} 
\frac{\omega^2}{c^2}\!=\!
\frac{\frac{1}{\epsilon_0}\int_I d\mathbf{r}
\nabla \mathcal{H}^{*}_z\!\cdot\!\nabla 
\mathcal{H}_z\!\!-\!\!\int_S d\mathbf{r}_S\int_S d\mathbf{r}'_S 
\mathcal{H}^{*}_z\!\left(\Sigma^m\!-\!\omega_0^2
\frac{\partial\Sigma^m}{\partial\omega_0^2}\right)\!\mathcal{H}_z}
{\int_I d\mathbf{r}\mathcal{H}^{*}_z\cdot \mathcal{H}_z+c^2\int_S 
d\mathbf{r}_S\int_S d\mathbf{r}'_S\mathcal{H}^{*}_z\frac{\partial
\Sigma^m}{\partial\omega_0^2}\mathcal{H}_z}\nonumber,
\end{equation}
where the magnetic embedding potential now satisfies
\begin{equation}\label{embm} 
\frac{1}{\epsilon_0}\frac{\partial H_z(\mathbf{r}_S)}
{\partial n_S}=\int_S d\mathbf{r}'_S\Sigma^m(\mathbf{r}_S,\mathbf{r}'_S)
\mathcal{H}_z(\mathbf{r}'_S).
\end{equation}

The matrix elements for determining the eigenmodes and Green function
are the same as in the case of $E$-polarization, apart from the matrix
element of the embedding potential, which with the definition
appropriate to the magnetic case, Eq.~(\ref{embm}), has an extra factor of
$1/\epsilon$,
\begin{eqnarray}
\int_S \!d\mathbf{r}_S\int_S \!d\mathbf{r}'_S F^*_i\Sigma^m F_j&=&
-\frac{2\pi r\kappa}{\epsilon(\omega_0)}\sum_m^{m_{\mbox{\tiny max}}}
\exp im(\phi_i-\phi_j)\nonumber \\&\times&\frac{J_m(k_i r)J_m(k_j r)}
{J_m(\kappa r)}J_m'(\kappa r).
\end{eqnarray}

Surface-plasmon polaritons occur with $H$-polarization, and the maximum
harmonic $m_{\mbox{\tiny max}}$ imposes a limit to the plasmon modes over each
cylinder, without which the density of states increases without limit
at the planar surface-plasmon frequency $\omega_p/\sqrt{2}$.
Controlling $m_{\mbox{\tiny max}}$ enables us to study the evolution
of the plasmon bands in detail. Of course there is a physical cut-off
for the plasmons due to Landau damping, though this is much larger
than the $m_{\mbox{\tiny max}}$ which we use. Unlike in scattering
theory, where a maximum multipole moment is also imposed, all higher
values of $m$ are included in our embedding, but with a different
boundary condition. In the case of $H$-polarization, the boundary
condition for $m>m_{\mbox{\tiny max}}$ is that the surface-normal
derivative of $H_z$ vanishes on the surface of the cylinders,
corresponding to the surface-parallel component of $\mathbf{E}$
vanishing.

Once again we evaluate the \emph{electric} spectral density, as the
plasmons are essentially electrostatic in nature. Using Maxwell's
equations, the integrated spectral density is given by
\begin{equation}
n_I(\omega)=\frac{2}{\pi\omega\epsilon_0 c^2}\sum_{ij}\mbox{Im}
\Gamma_{ij}\int_I d\mathbf{r}\nabla F^*_i\cdot\nabla F_j,
\end{equation}
where the $F_i$'s are the basis functions used to expand the magnetic
field between the cylinders.

\section{Results and discussion}
We have applied the embedding formalism described above to study a
square array of infinitely long Drude cylinders in vacuum. We use the
dimensionless reduced frequency $\tilde\omega=\omega a/2\pi c$ and
reduced wave vector $\tilde{\bf k}={\bf k}a/2\pi$, where $a$ is the
lattice constant. The reduced cylinder radius is defined as
$\tilde{r}=2\pi r/a$. 

First we provide a comparison of embedding with finite-difference
time-domain (FDTD) calculations by Ito and Sakoda,~\cite{takunori} in
which both space and time are discretized. They calculated the band
structure for $H$-polarization for Drude cylinders in vacuum with
reduced plasmon frequency $\tilde{\omega}_p=1$, and reduced radius
$\tilde{r}=1.885$. Results for this system calculated using embedding,
with 121 and 241 plane wave basis functions and $m_{\mbox{\tiny
max}}=6$ are shown in table 1, compared with the FDTD results, taken
from figure 5 of Ito and Sakoda.~\cite{takunori} We see first of all
that there is excellent agreement, except for band 2, where FDTD gives
an energy slightly higher than our (converged) results. We also note
that with only 121 plane waves the embedding method gives essentially
converged results, though the individual plasmon bands within the band
limits given in the tables do change slightly.
\begin{table}[h]
\begin{tabular}{r|c|c|c}
Band&121&241&FDTD\\ \hline
1&0.039&0.039&0.036\\
2&0.566&0.566&0.596\\
Plasmons&0.638 -- 0.711&0.638 -- 0.706&0.642 -- 0.718\\
3&1.088&1.086&1.082\\
4&1.161&1.161&1.164\\
5&1.184&1.184&1.182\\ \hline\hline
1&0.301&0.301&0.300\\
2&0.474&0.474&0.515\\
Plasmons&0.632 -- 0.720&0.632 -- 0.720&0.637 -- 0.728\\
3&0.921&0.921&0.927
\end{tabular}
\caption{Convergence test for $H$-polarised electromagnetic waves in a
square array of Drude cylinders in vacuum with $\tilde{\omega}_p=1$,
$\tilde{r}=1.885$. Upper table: frequencies of bands at
$\tilde{k}_x=0.05,\tilde{k}_y=0$ with 121 and 241 basis functions,
compared with FDTD results of Ito and Sakoda \cite{takunori}. Lower
table: the same for $\tilde{k}_x=0.5,\tilde{k}_y=0$}.
\vspace{0.2cm}
\end{table}

For studying the evolution of the plasmon bands, in the rest of this
paper we choose the reduced plasmon frequency $\tilde\omega_p=0.1$,
which for Al ($\omega_p\sim 15\,{\rm eV}$) corresponds to a lattice
constant of $a\sim 83\,$\AA. This value of $\tilde\omega_p$ gives
results which are universally applicable in the plasmonic
regime, where the radius of the cylinders is smaller than the wavelength of the
electromagnetic radiation. The reduced lifetime $\tau=2\pi c\tau/a$ is taken
as 1000,
broadening peaks and enabling us to plot densities of states easily as
a function of frequency.

\subsection{$E$-polarization}

\begin{figure}
\includegraphics[width=0.95\linewidth]{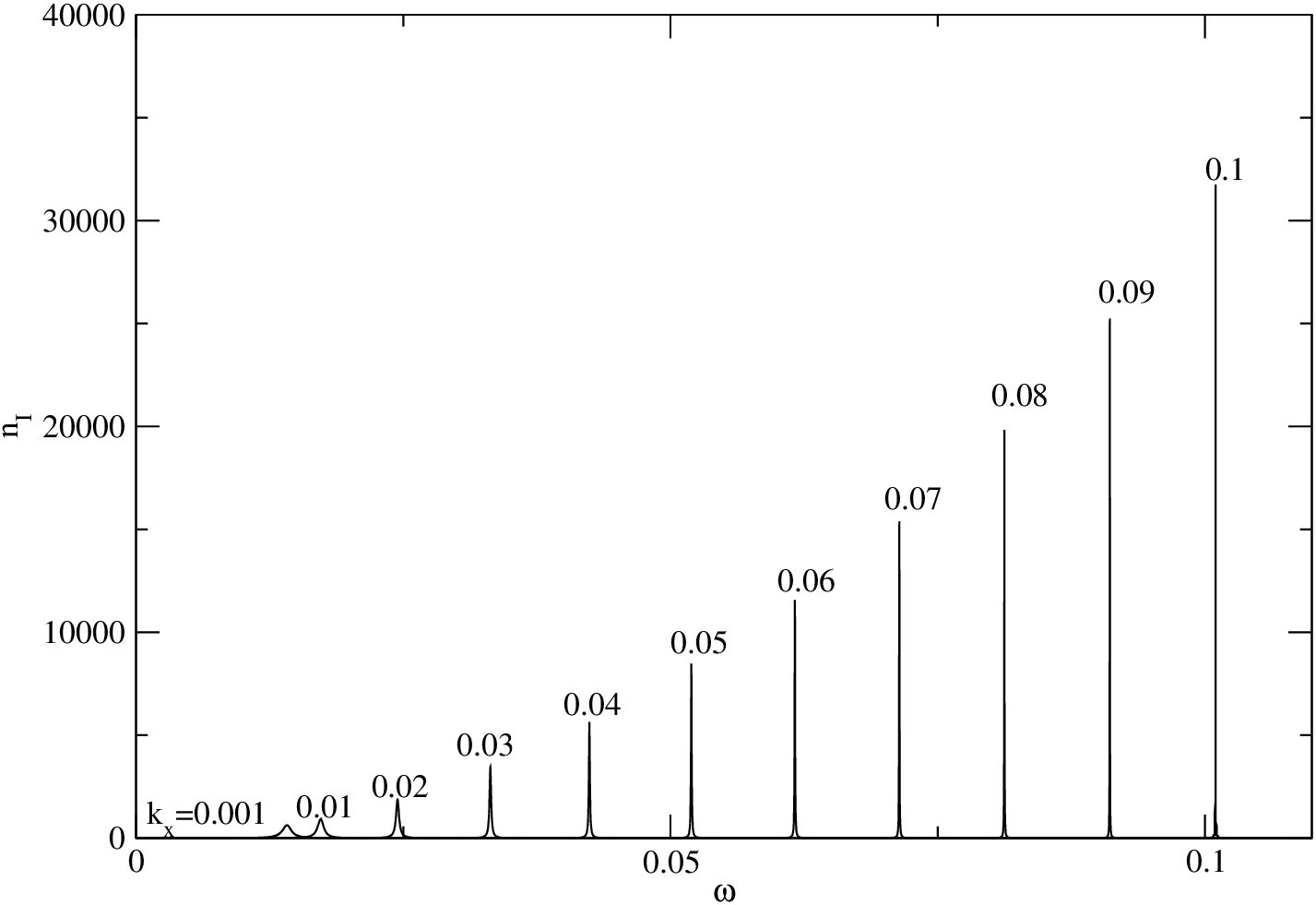}
\vspace{0.2cm}
\caption{Density of states $n_I(\omega)$ for $E$-polarized electromagnetic
waves in a square array of Drude cylinders in vacuum, with $\tilde{r}=0.5$
($f=0.02$). Different curves correspond to increasing values of $\tilde k_x$,
from 0.001 to 0.1 (with $\tilde k_y=0$).}
\vspace{0.5cm}
\label{fig1}
\end{figure}

\begin{figure}
\includegraphics[width=0.95\linewidth]{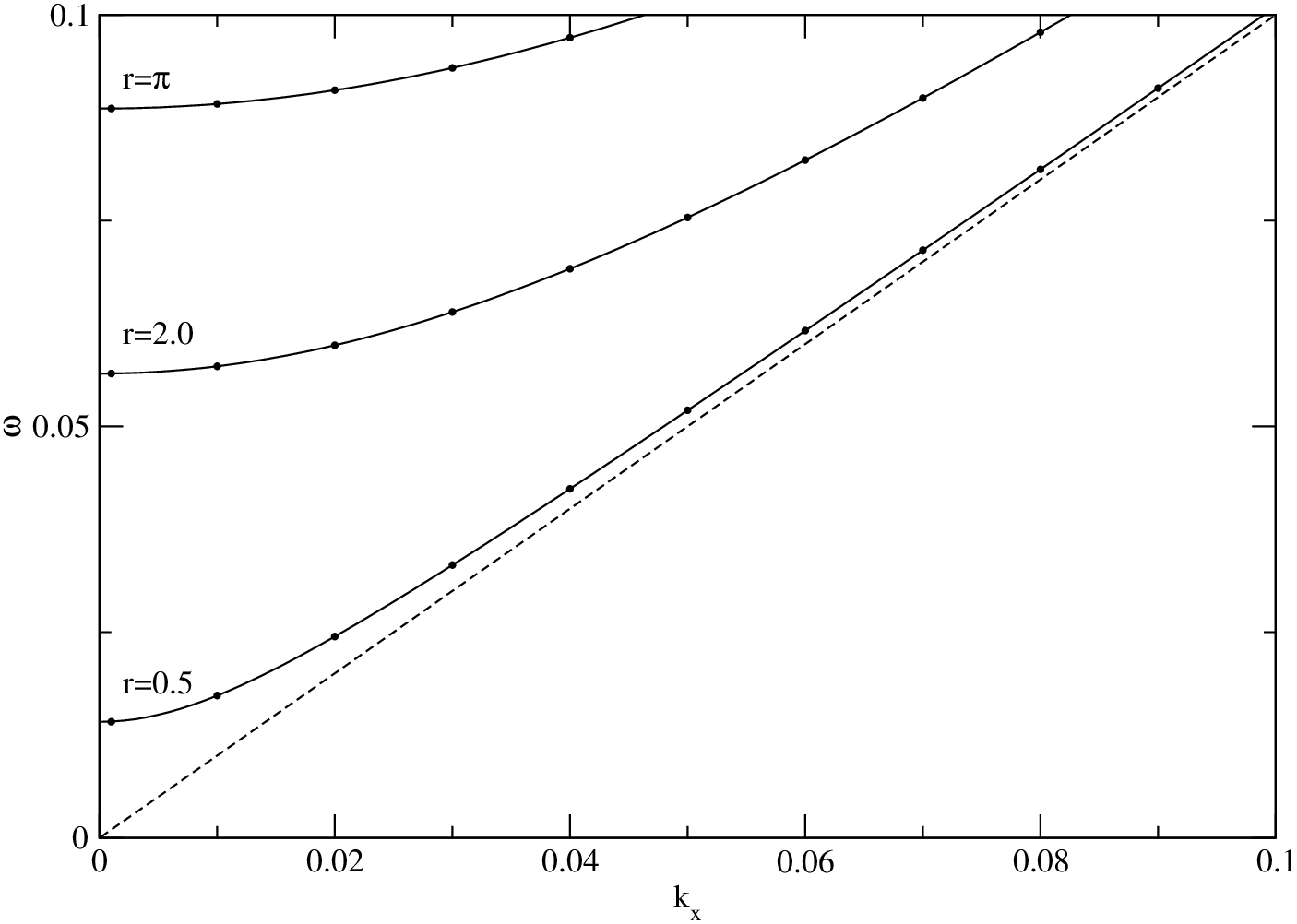}
\caption{$E$-polarization band structure as a function of
$\tilde{k}_x$ (with $\tilde{k}_y=0$), for cylinders of radius
$\tilde{r}=0.5,2.0$, and $\pi$ (touching cylinders); the corresponding filling
fractions $f$ are: 0.02, 0.32, and 0.79, respectively. The solid circles
represent the band structure that we have obtained from the peaks in our
calculated density of states $n_I(\omega)$. The solid lines correspond to a
homogeneous electron gas with
the reduced plasma frequency $\sqrt{f}\omega_p$. The dashed line is the light
line, $\omega=ck$.}
\label{fig2}
\vspace{0.5 cm}
\end{figure}

For $E$-polarized electromagnetic waves (with the electric field
parallel to the cylinder axis), our numerical results for the integrated
density of states $n_I(\omega)$ and the band structure $\omega(k)$ accurately
reproduce the results expected from Eqs.~(\ref{drude})-(\ref{epol1}) and,
therefore, the results reported in Ref.~\onlinecite{efros} for this
polarization. Figure~\ref{fig1} gives $n_I(\omega)$ for various values of
$k_x$ at a filling
fraction $f=0.02$.\cite{note} The corresponding
electromagnetic band structure $\omega(k)$, determined from the peak
frequencies, is shown in Fig.~\ref{fig2} (solid circles) for three values of
the filling fraction $f$. This figure shows that in the plasmonic r\'{e}gime
under consideration the optical response of the composite material is indeed
that of free electrons in a homogeneous electron gas, but with the reduced
plasma frequency $\sqrt{f}\omega_p$ (solid lines in Fig.~\ref{fig2}) expected
from Eqs.~(\ref{drude})-(\ref{epol1}).

The 2D periodicity of the composite material introduces
frequency gaps in the band structure, but these are remarkably
small. For cylinders with a reduced radius $\tilde{r}=2$,
corresponding to $f=0.32$, the first band gap at the centre
of the side of the 2D Brillouin zone is 0.0035 in
reduced units, and for touching cylinders with $f=0.79$ it is 0.0028.

\subsection{$H$-polarization}

\begin{figure}
\includegraphics[width=0.95\linewidth]{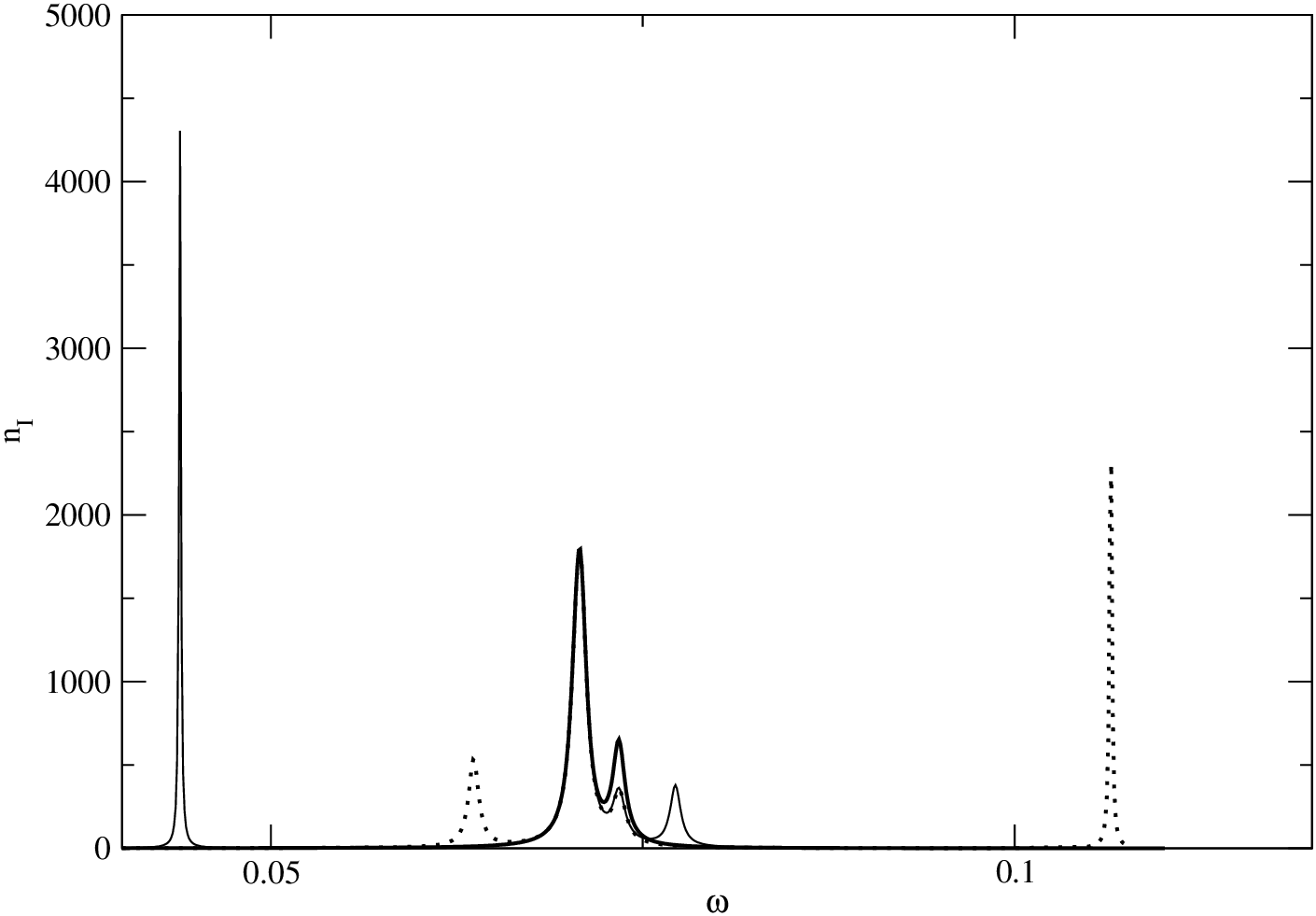}
\vspace{0.2cm}
\caption{Density of states $n_I(\omega)$ for $H$-polarized electromagnetic
waves in a square array of Drude cylinders in vacuum, with $\tilde{r}=1.0$
($f=0.08$), $m_{\mbox{\tiny max}}=4$, $\tilde{k}_y=0$, and varying
$\tilde{k}_x$: 0.001 (thick solid line), 0.05 (thin solid line), and
0.1 (dotted line).}
\label{fig3}
\end{figure}

For $H$-polarized electromagnetic waves (with the electric field
normal to the axes of the cylinders), our results are rather similar
to those found for a 3D lattice of metal spheres,\cite{spheres4} with
almost dispersionless surface plasmons on the cylinders and the light
line interacting with one of the two dipole plasmons. A significant
difference is that in the case of an isolated Drude metallic sphere,
the dispersionless Mie plasmons have frequencies given by
\begin{equation}\label{mie}
\omega_l=\omega_p\sqrt{\frac{l}{2l+1}},
\end{equation}
where $l$ is the multipole quantum number, whereas for an isolated
Drude cylinder \emph{all} the surface plasmons have the frequency
$\omega_p/\sqrt 2$, irrespective of quantum number $m$. This holds for
lattices with low filling fraction ($f\to 0$), and consequently low interaction
between the plasmon modes, resulting in a narrower range of plasmon
bands in the case of cylinders. For larger filling fractions, the
interactions between the cylinders dominate, and produce a wider
range of bands, similar to the spherical case.

\subsubsection{Density of states}

\begin{figure}
\includegraphics[width=0.95\linewidth]{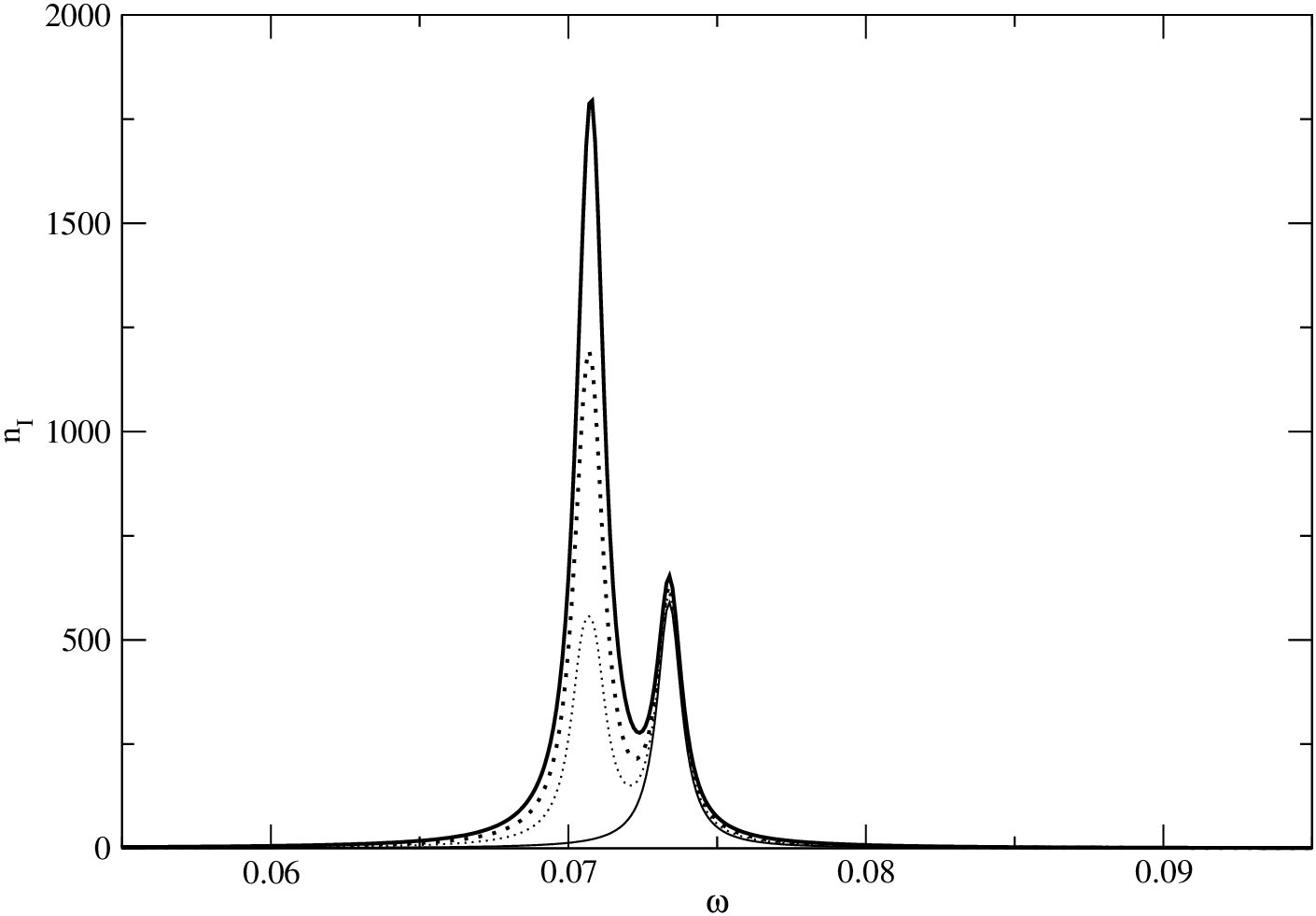}
\vspace{0.2cm}
\caption{Density of states $n_I(\omega)$ for $H$-polarized electromagnetic
waves in a square array of Drude cylinders in vacuum, with $\tilde{r}=1.0$
($f=0.08$), wave vector $\tilde{\mathbf{k}}=(0.001,0)$, and varying
$m_{\mbox{\tiny max}}$: 1 (thin solid line), 2 (stippled line), 3 (dotted
line), and 4 (thick solid line).}
\label{fig4}
\end{figure}

We start off by considering the integrated density of states. Figure~\ref{fig3}
shows $n_I(\omega)$ at three wave vectors, for cylinders with
$\tilde{r}=1$ corresponding to $f=0.08$. The cylindrical surface plasmons
constitute the {\it dispersionless} peaks near $\tilde\omega_p/\sqrt 2=0.071$,
with the light line dispersing through the plasmons and interacting with a
dipole mode.

In order to study the plasmons in more detail, we plot in Fig.~\ref{fig4}
the plasmon peaks for cylinders with $\tilde r=1$ (as in Fig.~\ref{fig3}) but
now at a small wave vector ($\tilde k_x=0.001$) and various values of
$m_{\mbox{\tiny max}}$: 1, 2, 3, and 4. We see that at this rather small
filling fraction ($f=0.08$), the plasmon structure is very narrow -- much
narrower than in the spherical case.\cite{spheres4} The effect of increasing
$m_{\mbox{\tiny max}}$ is simply to increase the height of the peak at
$\tilde{\omega}=0.071$, or $\tilde{\omega}_p/\sqrt 2$. In this sense,
the results do not converge as $m_{\mbox{\tiny max}}$ increases. But
this is physically correct -- the plasmon peak at
$\tilde{\omega}_p/\sqrt 2$ must increase without limit as the number
of allowed modes increases. However, the rest of the structure clearly
converges, as we see from Fig~\ref{fig4}.

\begin{figure}
\includegraphics[width=0.95\linewidth]{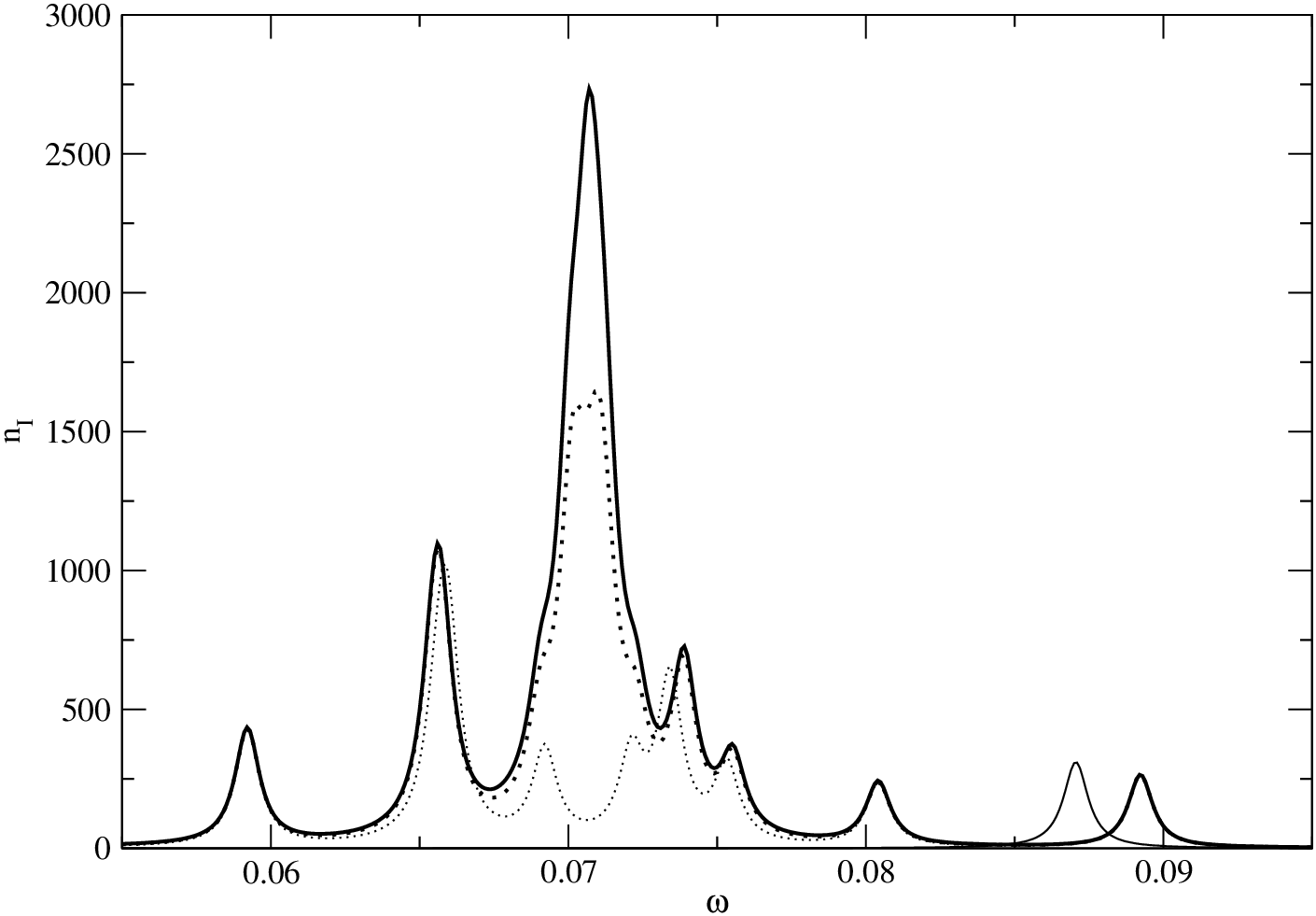}
\vspace{0.2cm}
\caption{Density of states $n_I(\omega)$ for $H$-polarized electromagnetic
waves in a square array of Drude cylinders in vacuum, with $\tilde{r}=2.62$
($f=0.55$), wave vector $\tilde{\mathbf{k}}=(0.001,0)$, and varying
$m_{\mbox{\tiny max}}$: 1 (thin solid line), 6 (stippled line), 10 (dotted
line), and 12 (thick solid line).}
\vspace{0.7cm}
\label{fig5}
\end{figure}

\begin{figure}
\includegraphics[width=0.95\linewidth]{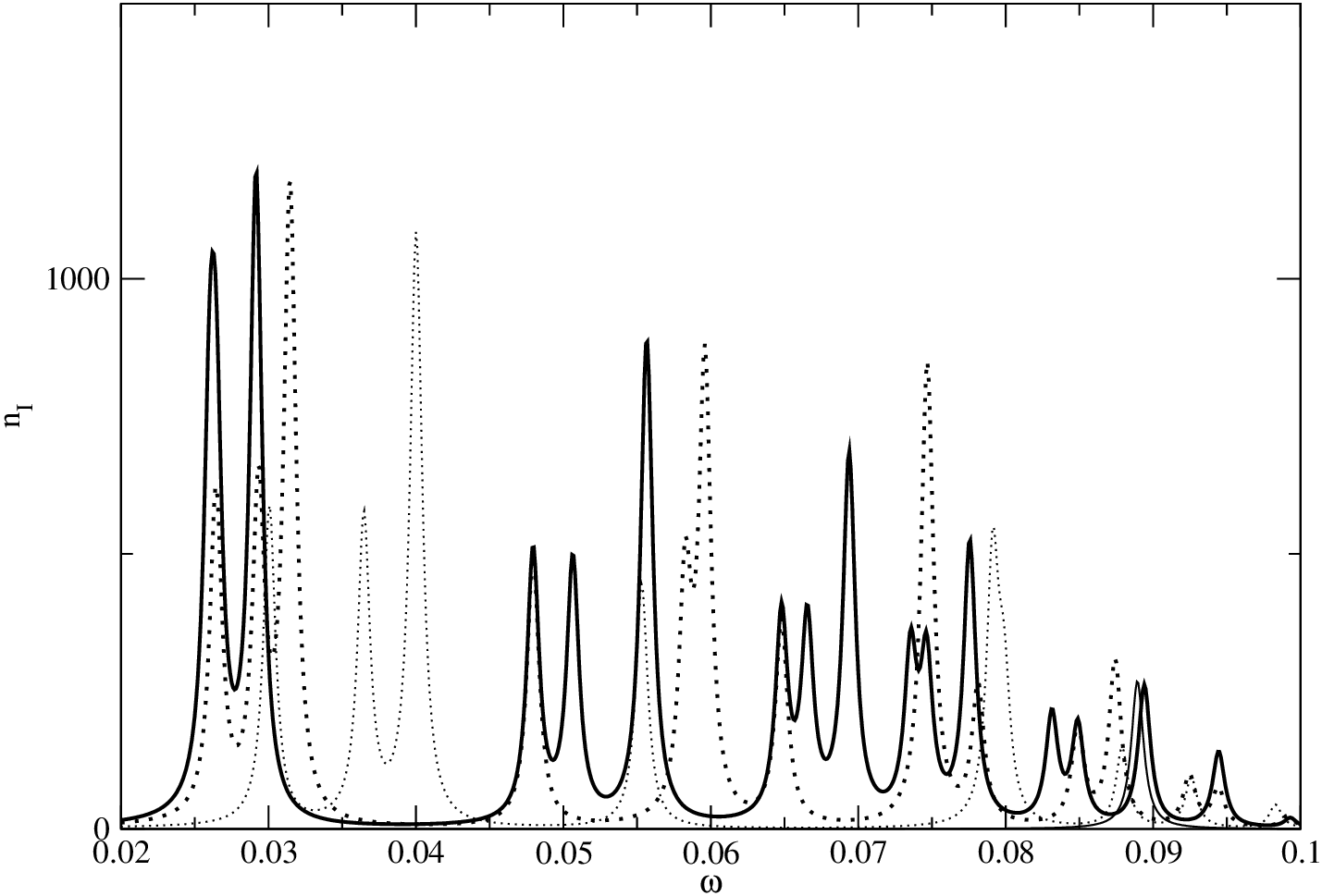}
\vspace{0.2cm}
\caption{Density of states $n_I(\omega)$ for $H$-polarized electromagnetic
waves in a square array of {\it touching} Drude cylinders in vacuum, with
$\tilde{r}=\pi$ ($f=0.79$), wave vector $\tilde{\mathbf{k}}=(0.001,0)$, and
varying $m_{\mbox{\tiny max}}$: 1 (thin solid line), 6 (stippled line), 10
(dotted line), and 12 (thick solid line).}
\label{fig6}
\end{figure}

This behaviour is confirmed for larger cylinder radii, when the
electrostatic interaction between the cylinders broadens the plasmon
structure. Figure~\ref{fig5} shows the density of states at a wave vector
close to zero ($k=0.001$, as in Fig.~\ref{fig4}), for a cylinder of radius
$\tilde{r}=2.62$ corresponding to a packing fraction of $f=0.55$. We see that
the wings of the plasmon structure still converge with $m_{\mbox{\tiny max}}$,
and once again increasing $m_{\mbox{\tiny max}}$ increases the height
of the central peak. For touching cylinders (Fig.~\ref{fig6}), the density of
states broadens to give plasmon structure between a rather low frequency and
the bulk plasmon frequency, $\tilde{\omega}_p=0.1$. In
this case, we cannot distinguish a central peak at the surface plasmon
frequency, and increasing $m_{\mbox{\tiny max}}$ alters the structure
within the entire plasmon range -- though the plasmon limits
converge. This is similar to the behaviour we found earlier for
touching spheres.~\cite{spheres4}

The size of basis set we need for the convergence of these
calculations is rather small, about 350 basis functions for the highest
$m_{\mbox{\tiny max}}$ under consideration: $m_{\mbox{\tiny max}}=4$ in
Figs.~\ref{fig3} and \ref{fig4}, and $m_{\mbox{\tiny max}}=12$ in
Figs.~\ref{fig5} and \ref{fig6}.

\subsubsection{Photonic band structure}

\begin{figure}
\includegraphics[width=0.95\linewidth]{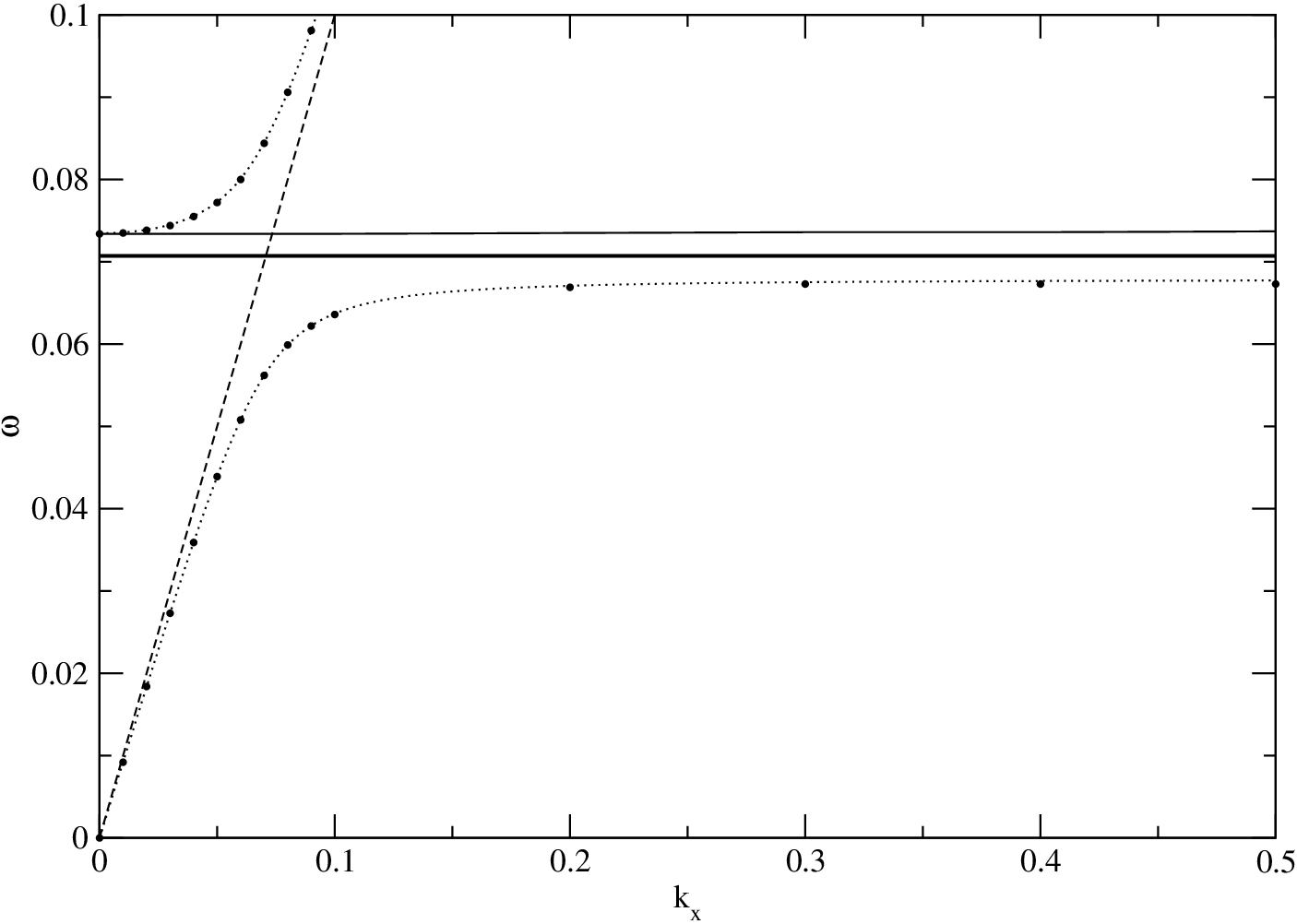}
\vspace{0.2cm}
\caption{$H$-polarization band structure as a function of
$\tilde{k}_x$ (with $\tilde{k}_y=0$, i.e., along the $\bar{\Gamma}\bar{M}$
direction), for cylinders of radius $\tilde{r}=1.0$ ($f=0.08$). The solid
lines and circles represent the band structure (converged in $m_{\mbox{\tiny
max}}$) that we have obtained from the peaks in our calculated density of
states $n_I(\omega)$. The dotted lines represent the $\omega(k)$ curves
derived from Eq.~(\ref{general}) with the MG effective
dielectric function $\epsilon_{\mbox{\tiny eff}}(\omega)$ of
Eq.~(\ref{mg}).}
\vspace{0.5cm}
\label{fig7}
\end{figure}

\begin{figure}
\includegraphics[width=0.95\linewidth]{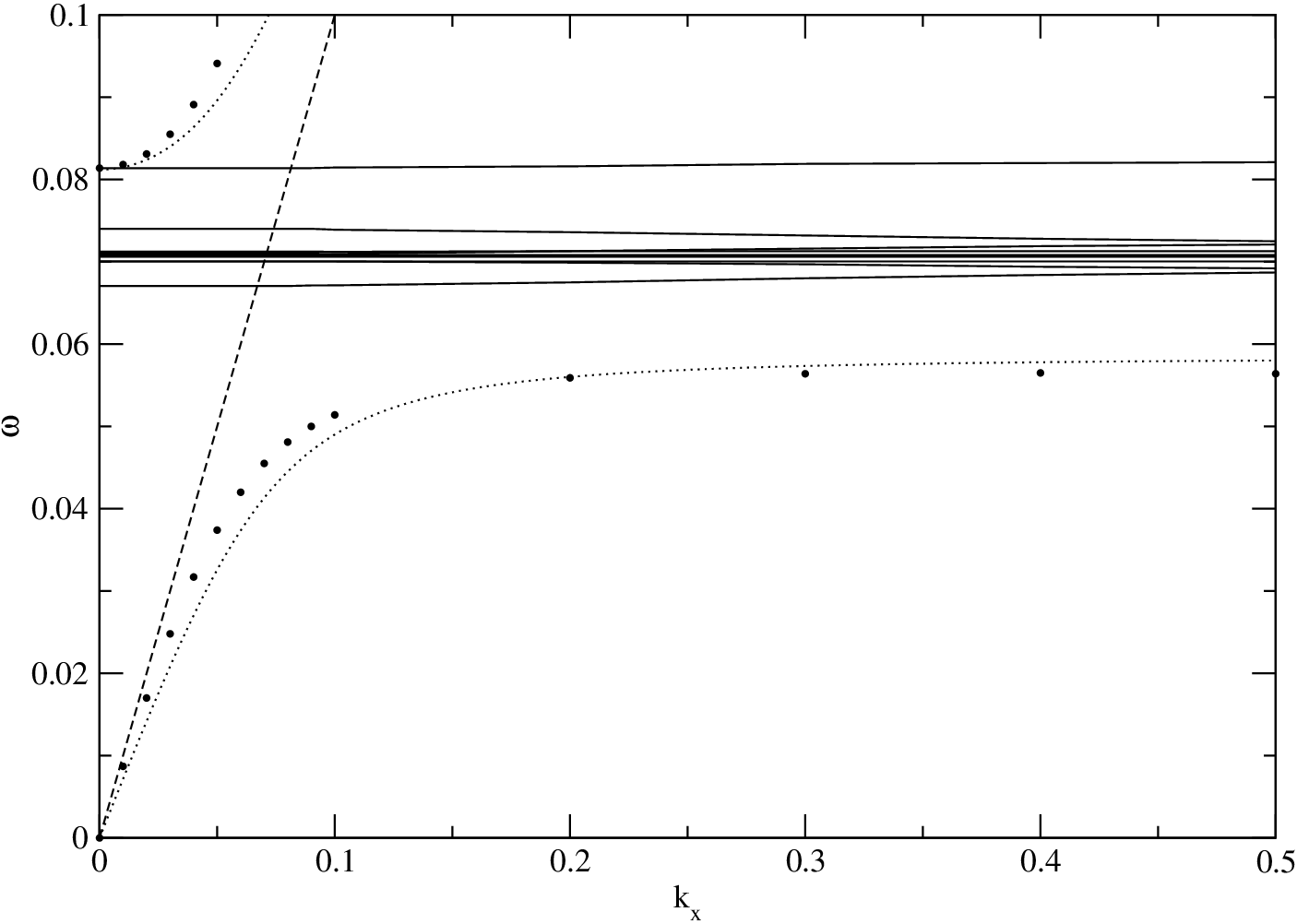}
\vspace{0.2cm}
\caption{As in Fig.~\ref{fig7}, but now for cylinders of radius $\tilde{r}=2.0$
($f=0.32$). This is converged in $m_{\mbox{\tiny max}}$.}
\vspace{0.5cm}
\label{fig8}
\end{figure}

\begin{figure}
\includegraphics[width=0.95\linewidth]{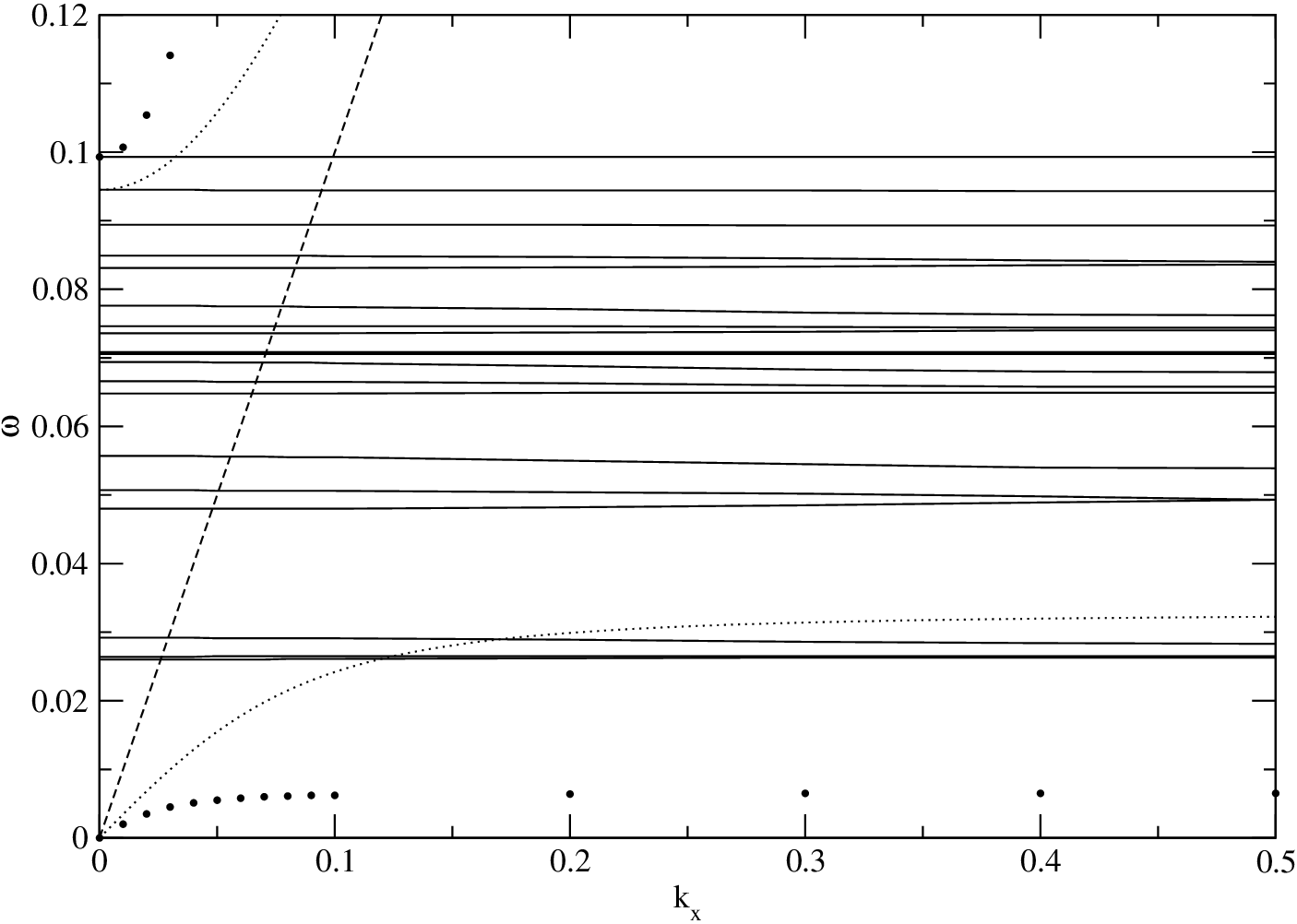}
\vspace{0.2cm}
\caption{As in Fig.~\ref{fig7}, but now for {\it touching} cylinders of radius
$\tilde{r}=\pi$ ($f=0.79$). $m_{\mbox{\tiny max}}=12$, and although the band
edges have converged, the individual bands have not.}
\label{fig9}
\end{figure}

The peaks in the integrated density of states $n_I(\omega)$ correspond to
normal modes of the system, and plotting the frequencies as a function of wave
vector we obtain the photonic band structures shown in
Figs.~\ref{fig7}--\ref{fig9}. Also plotted in these figures (dotted lines) are
the $\omega(k)$ curves derived from Eq.~(\ref{general}) with the MG effective
dielectric function $\epsilon_{\mbox{\tiny eff}}(\omega)$ of Eq.~(\ref{mg}).
The limit of the lower branch occurs at $\sqrt{(1-f)/2}\,\omega_p$,
where the MG $\epsilon_{\mbox{\tiny eff}}(\omega)$ has a pole, and the starting
frequency of the upper branch occurs at $\sqrt{(1+f)/2}\,\omega_p$, where
the MG $\epsilon_{\mbox{\tiny eff}}^{-1}(\omega)$ has a pole.~\cite{noteefros}
In
the dilute ($f\to 0$) limit, both the limit of the lower branch and the
starting frequency of the upper branch occur at the planar surface-plasmon
frequency $\omega_p/\sqrt 2$. 

The band structure in Fig.~\ref{fig7} is for a small cylinder radius,
$\tilde{r}=1$ (corresponding to a filling fraction of $f=0.08$), and
exhibits an infinite number of dispersionless plasmon modes at the
planar surface-plasmon frequency $\tilde\omega_p/\sqrt 2$. Light
(singly degenerate, as we are only dealing with $H$-polarization)
interacts with one of the dipole plasmons to give the light line
(represented by solid circles), which is reproduced almost exactly by
MG theory \cite{note2} (dashed lines). The second dipole plasmon gives
the almost dispersionless band joining the upper branch of the light
line at zero wave vector. The dipole-active light line branches span
the multipole plasmon branches here, all at $\tilde\omega_p/\sqrt 2$,
quite different from the case of a lattice of metal spheres, where it
is only for larger filling fractions that the light branches enclose
the multipole plasmons.

At the larger filling fraction, $f=0.32$, considered in
Fig.~\ref{fig8}, the dipole modes and the light branches move further
away from the planar surface-plasmon frequency, again following
closely the MG curves. The multipole plasmon modes of lowest order are
slightly broadened around $\tilde{\omega}_p/\sqrt 2$, but again there
is remarkably little dispersion. There is a considerable difference
between these results and those for the lattice of spheres with the
same radius, which shows a much wider spread of multipole frequencies,
and much more dispersion.
 
With touching cylinders (Fig.~\ref{fig9}), {\it all} multipole bands spread
out and MG breaks down badly. The frequency range of the multipole plasmon
bands is not quite as wide as in the case of touching spheres, and the lower
band edge seems to converge at $\tilde{\omega}\approx 0.026$ with
$m_{\mbox{\tiny max}}=12$. (In the case of spheres, the lower band edge
at $\tilde{\omega}\approx 0.02$ was still dropping at a maximum
multipole value of 12.) 

\section{Conclusions}

We have shown that as in the case of a lattice of metallic
spheres\cite{spheres4,ingles1} the embedding method provides a very
economical method of calculating electromagnetic waves in a lattice of
metallic cylinders,\cite{notelast} which unlike other methods allows to
describe accurately plasmon modes in the whole range of filling fractions from
the dilute limit to the case of touching wires. For the case of
$H$-polarization, where surface
plasmons dominate the normal modes, the results are quite similar to
those obtained for a lattice of metallic spheres, except that the frequency
range of the multipole modes is narrower. For $E$-polarization, the
system behaves almost exactly like a dilute electron gas, with a
reduced plasmon frequency and only minute band gaps at the Brillouin
zone boundaries.

The method and results presented here are for the case of the
wave vector lying in the plane perpendicular to the cylinders.
We are also considering the general case of arbitrary wave vector, and
this will be the subject of a subsequent paper.

\acknowledgments

J.M.P. acknowledges partial support by the University of the Basque Country,
the Basque Unibertsitate eta Ikerketa Saila, the Spanish Ministerio de
Educaci\'on y Ciencia, and the EC 6th framework Network of Excellence
NANOQUANTA (Grant No. NMP4-CT-2004-500198).









    

\begin{thebibliography}{100}
\bibitem{noter} The effective dielectric tensor of a lattice of metallic
cylinders was reported in Ref.~\onlinecite{efros}, at all wavevectors of the
electromagnetic waves, in the dilute ($f\to 0$) limit. When the wavevector is
perpendicular to the axis of the cylinders, their Eq.~(23) for $s$-polarization
yields our Eq.~(\ref{epol1}).
\bibitem{note} For a 2D square array of cylinders of radius $r$ and lattice
constant $a$, the filling fraction is $f=\tilde r^2/4\pi$, $\tilde r$ being
the reduced radius: $\tilde r=2\pi r/a$.
\bibitem{noteefros} This is in contrast with the less accurate results reported
in Ref.~\onlinecite{efros} for $p$-polarization, obtained from Eq.~(\ref{mg0})
[their Eq.~(27)], where the limit of the lower branch and the starting
frequency of the upper branch occur at $\omega_p/\sqrt{2}$ and
$(1+f)\omega_p/\sqrt{2}$, respectively.
\bibitem{notelast} Calculations of band structures of a lattice of spheres were
compared in Ref.~\onlinecite{ingles1} with the results obtained with other
plane-wave methods, showing that embedding needed far fewer plane waves, as
expected.
\end{thebibliography}
\end{document}